\documentclass{aa}
\usepackage{psfig}
\def \sax {BeppoSAX}
\def \degmark{^\circ}
\def \nh {N${\rm _H}$}
\def \ergsec{\hbox{erg s$^{-1}$}}
\def \hcm {\hbox {\ifmmode $ atom cm$^{-2}\else atom cm$^{-2}$\fi}}
\def \arcmin {\hbox{$^\prime$}}
\def \arcsec {\hbox{$^{\prime\prime}$}}

\def\approxgt{\mathrel{\hbox{\rlap{\lower.55ex \hbox {$\sim$}}
        \kern-.3em \raise.4ex \hbox{$>$}}}}
\def\approxlt{\mathrel{\hbox{\rlap{\lower.55ex \hbox {$\sim$}}
        \kern-.3em \raise.4ex \hbox{$<$}}}}
\newcommand{\mc}{\multicolumn}
\begin{document}

\thesaurus{23(03.09.1; 03.13.2; 03:20.3; 12.04.2; 13.25.3)}  

\title{\sax\ LECS background subtraction techniques}

\author{A.N. Parmar\inst{1}
        \and T. Oosterbroek\inst{1}
        \and A. Orr\inst{1}
        \and M. Guainazzi\inst{1}
        \and N. Shane\inst{1,2}
        \and M.J. Freyberg\inst{3}
        \and D. Ricci\inst{4}
        \and A. Malizia\inst{4,5}
}
\offprints{A.N.Parmar (aparmar@astro.estec. esa.nl)}

\institute{
        Astrophysics Division, Space Science Department of ESA, ESTEC,
              Postbus 299, 2200 AG Noordwijk, The Netherlands
   \and
        Pembroke College, University of Oxford, St Aldates, 
        Oxford, OX1 1DW, UK
   \and
        Max-Planck-Institut f\"ur Extraterrestrische Physik, D-85740 
        Garching bei M\"unchen, Germany
   \and
        BeppoSAX Science Data Center, Nuova Telespazio, via Corcolle 19
        I-00131 Roma, Italy
   \and
        Department of Physics and Astronomy, Southampton University,
        SO17~1BJ, UK
}
\date{Received 16 November 1998; accepted 2 February 1999}

\maketitle

\markboth{LECS background subtraction}{LECS background subtraction}

\begin{abstract}
We present 3 methods for the subtraction of non-cosmic and
unresolved cosmic backgrounds observed by  
the Low-Energy Concentrator Spectrometer (LECS) on-board \sax.
Removal of these backgrounds allows a more accurate modeling of the
spectral data from point and small-scale extended sources.
At high ($>$$\vert 25 \vert \degmark$) galactic latitudes, subtraction
using a standard background spectrum works well.
At low galactic latitudes, or in complex regions of the X-ray sky,
two alternative methods are presented. The first uses counts obtained from
two semi-annuli near the outside of the LECS field of view
to estimate the background at the source location. The second method uses
ROSAT Position Sensitive Proportional Counter 
(PSPC) all-sky survey data
to estimate the LECS background spectrum for a given pointing position.
A comparison of the results from these methods provides an estimate
of the systematic uncertainties.
For high galactic latitude fields, all 3 methods give 3$\sigma$ confidence
uncertainties of $<$$0.9 \times 10^{-3}$~count~s$^{-1}$ (0.1--10~keV),
or $<$$1.5 \times 10^{-3}$~count~s$^{-1}$ (0.1--2~keV). 
These correspond to 0.1--2.0~keV fluxes of
0.7--1.8 and 0.5--$1.1 \times 10^{-13}$~erg~cm$^{-2}$~s$^{-1}$
for a power-law spectrum with a photon index of 2
and photoelectric absorption of ${\rm 3 \times
10^{20}}$ and ${\rm 3 \times 10^{21}}$~atom~cm$^{-2}$, respectively.
At low galactic latitudes, or in complex regions of the X-ray sky,
the uncertainties are a factor $\sim$2.5 higher.
\end{abstract}


\section{Introduction}

The Low-Energy Concentrator Spectrometer
(LECS) (\cite{parmar97b}) is an imaging gas scintillation proportional
counter on-board the Italian-Dutch \sax\ X-ray astronomy mission.
\sax\ (\cite{boella97a}) was launched into a 600~km 
3\fdg9 inclination orbit on 1996 April 30. The LECS is one of four
co-aligned narrow field instruments (NFIs). The other NFIs
are the imaging Medium-Energy Concentrator
Spectrometer (MECS; 1.3--10~keV; \cite{boella97b}),
the High Pressure Gas Scintillation Proportional Counter
(HPGSPC; 5--120~keV; \cite{manzo97}) and the Phoswich
Detection System (PDS; 15--300~keV; \cite{frontera97}).
In addition, the payload includes two wide field cameras 
(WFC; 2--30~keV; \cite{jager97}) which observe in directions 
perpendicular to the NFI. The MECS background is discussed in
\cite{chiappetti98}.

Accurate background subtraction is important for virtually all 
X-ray observations and especially for studies of extended
and diffuse objects.
In this {\it paper} we discuss background subtraction techniques 
appropriate to the LECS.
Background subtraction is complicated in the
LECS because (1) the wings of the mirror point spread
function distribute a significant
fraction of hard X-rays over the entire field of view (FOV), (2) 
the broad detector response spreads a fraction
of low-energy X-rays over the entire FOV, (3) the off-axis response
of the mirrors is complicated and single scattered X-rays from outside
the direct FOV can reach the focal plane (\cite{conti94}),
and (4) at low-energies the X-ray sky has a great deal of spatial structure
(see e.g., \cite{snowden95}, \cite{snowden97}).
This spatial structure has a complex energy dependence. 
Below 0.28~keV there is a galactic plane-to-pole variation
of a factor $\sim$3, while at energies between 0.5 and
2.0~keV the North Polar Spur and the galactic bulge dominate.
For energies $>$0.5~keV and ${\rm 50\degmark < l_{II} < 290\degmark}$ 
the X-ray sky is relatively smooth with few discrete features
visible.

\section{The LECS}

\begin{figure}
  \centerline{\psfig{figure=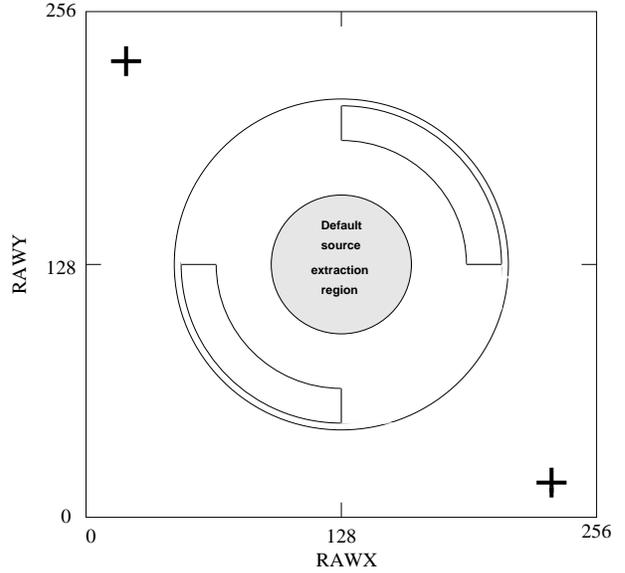,width=8.0cm,angle=0}}
  \caption[]{The LECS FOV drawn to scale. The outer circle indicates 
             the 37\arcmin\
             diameter sky FOV which contains the standard 16\arcmin\ 
             diameter source extraction region (shaded) and the two
             background semi-annuli.
             The positions of the two $^{55}$Fe radioactive sources are 
             shown as crosses}
\label{fig:cartoon}
\end{figure}

The method of operation of the LECS is similar to that of 
conventional gas scintillation proportional counters.
An X-ray that passes through the ultra-thin (1.25~$\mu$m) entrance
window and is absorbed in the 11.4~cm diameter Xe filled gas cell liberates
a cloud of electrons. A uniform electric field between the entrance window,
kept at $-$20~kV, and a grounded grid causes scintillation as the
electrons travel towards the grid. The UV light from these scintillations
is detected by a multi-anode photo-multiplier tube (PMT).
The LECS is also sensitive to cosmic rays since these can
ionize the counter gas and leave tracks through the cell.
The 2~cm diameter gas cell entrance window is supported by a tungsten
strongback and fine grid.
The strongback consists of a square grid of tungsten bars with a 
separation of 4\arcmin. Each strongback square is further divided into
8 by 8 pixels by a fine grid. The overall effect of the strongback
and fine grid is to reduce the X-ray transmission by between 20--40\%,
depending on position within the FOV. The LECS includes
two $^{55}$Fe radioactive sources which constantly illuminate regions
of the detector outside the sky FOV to provide monitoring of the
instrument performance (see Fig.~\ref{fig:cartoon}). 

The LECS mirror system consists of 30 nested, 
Au coated mirrors with a double cone approximation to the 
Wolter~I geometry. The mirror focal length is 185~cm and the geometric
area 124~cm$^2$ (\cite{conti94}, \cite{conti96}). 
The off-axis behavior of the mirrors
is complicated. Conti et al. (1994) demonstrate that X-rays within a cone
of solid angle 2$\degmark$ can reach the focal plane after reflection
off only one of the mirror surfaces. 
As part of the \sax\ Science Verification
Phase, observations were performed with the Crab Nebula just
outside the LECS FOV. These show that at an offset of 
60\arcmin, the Crab Nebula is visible
as an extended emission region offset in the direction of the source,
with an intensity $(5.2 \pm ^{1.8} _{1.4}) \times 10^{-4}$ of that 
on-axis. The spectrum of the offset emission (a power-law with a
photon index, $\alpha$, of $2.2 \pm 0.2$) is consistent with that
recorded from the Crab Nebula on-axis ($\alpha = 2.1$).

The LECS is only operated during satellite night time. Although
this reduces the observing efficiency considerably, it means 
that that any contribution to the background from 
scattered solar X-rays, e.g., as seen by the ROSAT Position Sensitive
Proportional Counter (PSPC) when the Sun-Earth-satellite angle is
$<$120$\degmark$ (e.g., \cite{snowden93}), is negligible.
The \sax\ observing schedule is designed to optimize the LECS
observing efficiency. This means that, in general, only time constrained
or Target of Opportunity observations include long intervals of dark Earth
pointing. 
The LECS is not operated when the satellite passes 
close to the South Atlantic Anomaly (SAA). 
Due to the low \sax\ orbital inclination SAA passages result in between 5 and
12~min of data being lost per 96~min satellite orbit.
In order to maximize the observing efficiency, many \sax\ NFI
observations are made with a Sun-satellite-target angle close to 120$\degmark$,
which is as far from the solar direction as it is possible to operate.

\begin{figure}
  \centerline{\psfig{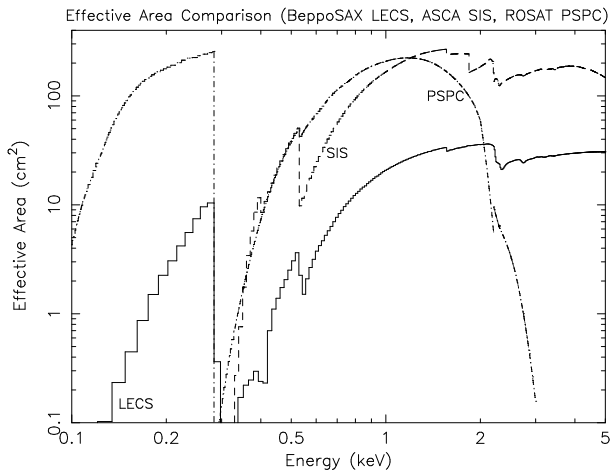}}
  \caption[]{A comparison of the LECS (solid line), ASCA SIS (dashed line)
and ROSAT PSPC (Dashed-dotted line) effective areas in the energy 
range 0.1--5.0 keV}
\label{fig:plot_effic3}
\end{figure}

The LECS energy resolution is 32\% full-width half-maximum
(FWHM) at 0.28~keV and 8.8\% FWHM at 6~keV.
The FOV is circular
with a diameter of 37\arcmin. The position resolution 
corresponds to 90\% encircled energy within a radius of
2\farcm5 at 1.5~keV. This is a factor
$\sim$4 worse than that of the PSPC and comparable to that of the 
ASCA Gas Imaging Spectrometer (GIS). 
Fig.~\ref{fig:plot_effic3} shows the on-axis LECS effective area. 
A key scientific goal of the LECS is to study sources in the
energy band below the instrument's C edge at 0.28~keV. This typically 
means that the absorption to a source must be 
$\approxlt$$3 \times 10^{21}$~\hcm.
The prime advantage of the LECS, compared to previous high throughput
X-ray detectors, is its good low-energy spectral resolution and its low
background afforded by the imaging characteristics.
The 0.1--2.0~keV energy resolution is a factor $\sim$2.4 
better that that of the PSPC,
while the effective area is between a factor $\sim$20 and 5 lower at 
0.28~keV and
1.5~keV, respectively. The energy resolution is similar to that of the
GIS in the overlapping energy range, and comparable to that of the
ASCA Solid-state Imaging Spectrometer at energies of $\sim$0.5~keV. 

\section{Data selection and processing}
\label{sect:data_sel}

In order to study the properties of the LECS background, data were extracted
from 350 \sax\ Final Observation Tapes, corresponding
to 170 individual observations. 
These data were processed using the standard LECS processing pipeline
described below and cleaned event lists for the intervals when
the instrument was observing both the dark Earth and the
sky in the nominal operating configuration extracted.
Observations with $<$100 counts were excluded. This left
total sky and dark Earth exposures of 2847~ks and 498.6~ks, respectively.

Table~\ref{tab:telemetry} gives the contents of the LECS telemetry
for each detected event. The standard LECS data processing package 
({\sc saxledas 1.8.0}) performs the following steps to produce cleaned,
linearized event lists:
\begin{enumerate}
\item{}Conversion of RAWX and RAWY into linear coordinates referred to
as DETX and DETY using the algorithm given in Parmar et al. (1997b).
\item{}Determination of the overall instrument gain (which is temperature
dependent), using the mean energy of the two calibration sources as
a reference. 
\item{} Conversion of Pulse height channels to Pulse Invariant
(PI) channels which have a fixed relation to energy. This
conversion takes into
account the overall gain and a FOV position dependent gain correction.
\item{}Selection of intervals with nominal instrument settings (such
as high voltage settings), and separation into events that are $>$4$\degmark$
from the Earth's limb (taken to be good on-target intervals), and those
where the dark Earth occults the FOV.
\item{} Removal of background events using the VETO and BL signals. The
VETO signal is the ratio between the light measured by the central PMT anode
and the sum of that measured by the surrounding anodes. 
It therefore has a high value
for on-axis X-ray sources and lower values for events occurring further 
off-axis. In {\sc saxledas 1.8.0} a sliding cell VETO window is applied
to the data. The BL signal is a measure of the duration of the light
flash produced following absorption in the gas cell. It depends on the
penetration depth of an event into the cell, which is energy dependent.
In {\sc saxledas 1.8.0} events are selected if their PI and BL values
are within a set of 62 pre-defined boxes.
\end{enumerate}

The standard LECS data analysis technique is to extract counts
within a radius of 8$'$ (corresponding to 35 RAW pixels; see 
Fig.~\ref{fig:cartoon}) 
from the source centroid using the cleaned event lists produced
by {\sc saxledas}. This large radius ensures
that 95\% of the 0.28~keV X-rays are included in the analysis. 
For absorbed, or very faint sources, extraction radii of 4\arcmin\ or
6\arcmin\ may be appropriate.
The \sax\ pointing accuracy is such that most target sources are located
within $\sim$2$'$ of the nominal position of RAWX, RAWY = 131, 124.

\begin{table}
\caption[]{LECS Telemetry Contents}
\begin{flushleft}
\begin{tabular}{lll}
\hline\noalign{\smallskip}
Parameter & Range & Comment \\
\noalign{\smallskip\hrule\smallskip}
RAWX   & 0--255 & Event X coordinate \\
RAWY   & 0--255 & Event Y coordinate \\
TIME   &  \dots & Event Time (units of 16$\mu$s) \\
PHA    & 0--1023 & Event pulse height (energy) \\
VETO   & 0--255 & Event light distribution \\
BL     & 0--255 & Event burstlength \\
\noalign{\smallskip}
\hline
\end{tabular}
\end{flushleft}
\label{tab:telemetry}
\end{table}

\section{Standard background}
\label{sect:standard_background}

The usual method for LECS background subtraction is to 
subtract the counts obtained in the same detector region using 
a standard background exposure. This consists of the sum of the
blank field exposures listed in Table~\ref{tab:blank_fields}.
The LECS data for these fields were processed using {\sc saxledas 1.8.0}.
No point sources are present in the individual fields with 0.1--2.0~keV
fluxes $>$$1.7 \times 10^{-13}$~\ergsec, or 2--10~keV fluxes of
$>$$9.7 \times 10^{-14}$~\ergsec.
Searches were also made of X-ray
catalogs such as the ROSAT Bright Source Catalog (\cite{voges96}) 
to check for the presence of contaminating point sources in,
and a couple of degrees outside the FOV. 
Note that the Gal cent-2 and -3 fields include 
the complex region of sky around the Draco nebula 
(e.g., \cite{kerp94}; \cite{moritz98}).
A 22.1~ks \sax\ background exposure on 1996~August~11 at 
RA${\rm = 14^h 42^m 27\fs1}$, Dec$= +19\degmark 44\arcmin 10\arcsec$
was not included in the standard background since it 
is in the direction of the North Polar Spur.
Examination of the LECS spectrum showed that it has significantly more 
low-energy flux than the fields listed in Table~\ref{tab:blank_fields}.
The remaining exposure time is 558.6~ks.

\begin{figure*}
  \centerline{\hbox{
              \psfig{figure=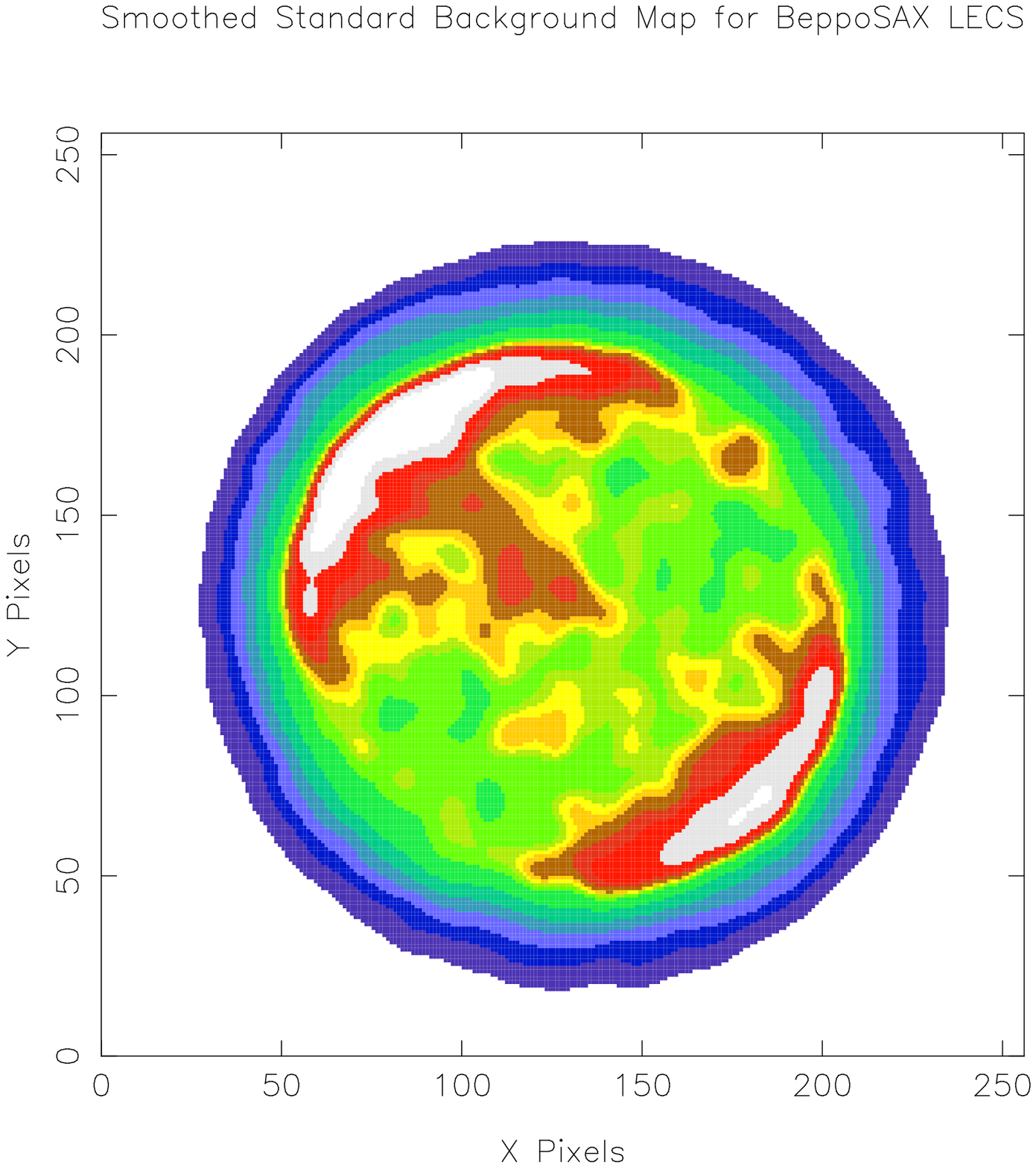,width=9.0cm,angle=-90}
              \psfig{figure=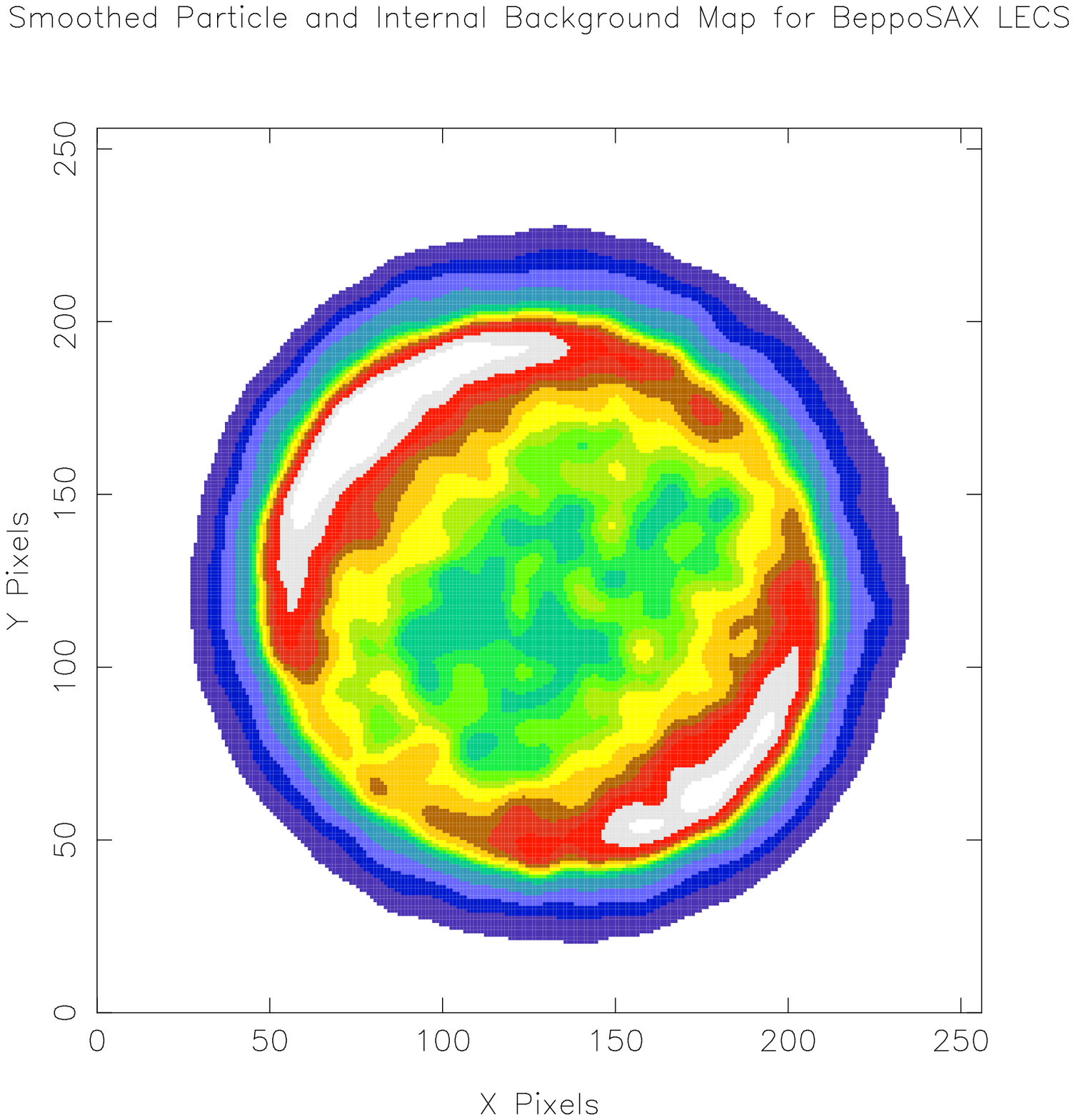,width=9.0cm,angle=-90}
  }}
  \caption[]{The dependence of the LECS standard background (left panel) and
             NXB (right panel) on
             RAWX, RAWY position within the FOV. Gaussian smoothing
             filters with a $\sigma$ of 4 pixels have been applied}
\label{fig:image}
\end{figure*}

The standard background 0.1--10.0~keV count rate in the 
central 8\arcmin\ radius source extraction region is 
$1.0 \times 10^{-5}$~s$^{-1}$~keV$^{-1}$~arcmin$^{-2}$.
The dependence of the standard background count rate
on position within the FOV is
shown in the left panel of Fig.~\ref{fig:image}. 
This image has been smoothed using
a Gaussian filter with a $\sigma$ of 4 RAW pixels. The difference in
intensity between the most and least intense regions
is a factor 1.8. 
The X-ray background appears brighter close to the center of the
image due primarily to mirror vignetting. Note that
the LECS mirror axis is offset by 3\arcmin\
from the center of the FOV due to an alignment error during integration.
The two bright arcs are located
radially inwards from the $^{55}$Fe calibration sources and
result from 5.9~keV characteristic 
X-rays that are absorbed some distance
from their origin. 

Fig.~\ref{fig:skybg_spec} illustrates the overall shape of the 
standard background spectrum obtained in the central 8\arcmin\ of the
FOV. The Non X-ray
background (NXB) obtained from the same region of the detector is also shown
(see Sect.~\ref{sect:nxb}).
The difference between the two is the contribution
of the cosmic X-ray background (CXB), spectral fits to which
are presented in a companion {\it paper} (\cite{parmar99}).

\begin{table*}
\caption[]{LECS observations used to create the
standard background spectrum. A target name including ``sec'' or 
``secondary'' refers to the prime WFC target. N is the number of
individual pointings. ${\rm N_{gal}}$
is the line of sight absorption in units of $10^{20}$~\hcm\ 
(\cite{dickey90})}
\begin{flushleft}
\begin{tabular}{llllllrlrl}
\hline\noalign{\smallskip}
       Target & \mc{2}{c}{Observation} & N & \mc{2}{c}{Position (J2000)} 
       & ${\rm l_{II}}$ \hfil
       & \hfil ${\rm b_{II}}$ & Exp. & ${\rm N_{gal}}$\\
       & \hfil Start  & \hfil Stop&& \hfil RA & \hfil Dec & $(\degmark)$ \hfil 
       & \hfil $(\degmark)$ & (ks) & \\
       & (yr~mn~day) & (yr~mn~day) & & \hfil (h~m~s) & \hfil ($\degmark\ 
       \arcmin\ \arcsec$) & & & & \\
\noalign{\smallskip\hrule\smallskip}
Polaris Region   & 1996 Jul 01 & 1997 May 23 & 6 &02 31 42.0 & +89 15 47 & 
123.3 & +26.5 & 69.7 & 7.0 \\
Gal cent-2 (sec) & 1996 Aug 23 & 1996 Aug 27 & 1 &16 51 19.5 & +60 11 49 & 
89.8 & +38.1 & 82.3 & 2.1\\
Gal cent-3 (sec) & 1996 Aug 27 & 1996 Aug 31 & 1 &16 35 10.7 & +59 46 30 & 
89.9 & +40.2 & 79.9 & 2.0\\
SGR-A (sec)      & 1996 Oct 10 & 1996 Oct 12 & 1 &06 12 22.7 & $-$60 59 04 & 
270.1 & $-$28.2 & 43.3 & 4.2\\
Secondary Target (S1)& 1997 Mar 18 & 1997 Mar 25 & 1 &17 56 46.1 & +61 11 45 & 
90.2 & +30.1 & 94.2 & 3.4\\
Secondary Target (S2)& 1997 Apr 13 & 1997 Apr 15 & 2 &18 18 20.5 & +60 58 42 & 
90.2 & +27.4 & 26.6 & 3.8\\
SDC Target       & 1997 Dec 13 & 1997 Dec 14 & 1 &23 07 53.5 & +08 50 06 &
84.4 & $-$46.1 & 13.9 & 4.7 \\
Secondary Target (S4)& 1998 Mar 10 & 1998 Mar 22 & 2 &05 52 07.9 
& $-$61 05 35 & 270.0 & $-$30.6 & 46.7 & 5.3 \\
Secondary Target (S5)& 1998 Aug 22 & 1998 Oct 01 & 4 &17 52 07.4 & +61 01 01 &
89.9  & +30.6 & 102.2 & 3.5 \\
\noalign{\smallskip}
\hline
\end{tabular}
\end{flushleft}
\label{tab:blank_fields}
\end{table*}

\section{Non X-ray background (NXB)}
\label{sect:nxb}

\begin{figure*}
  \centerline{\psfig{figure=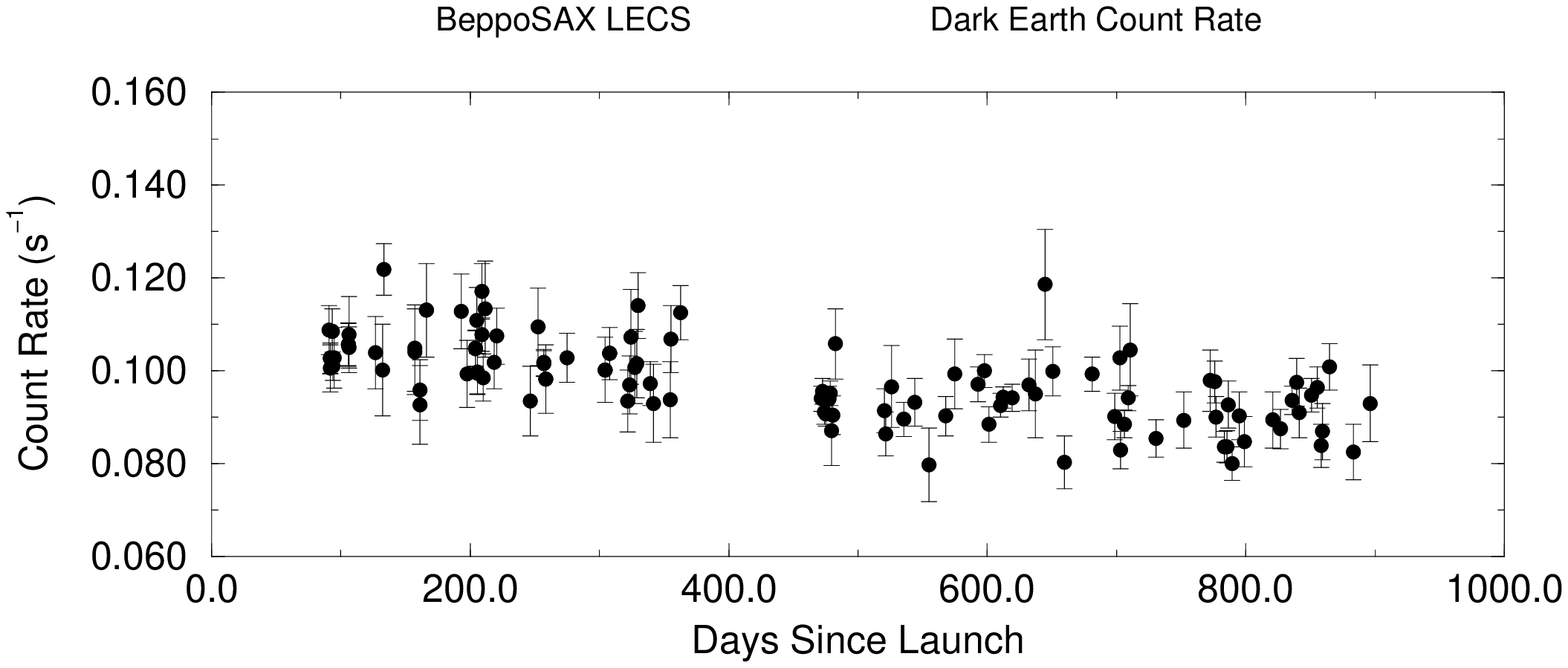,width=18.5cm,angle=-90}}
  \vspace{-0.7cm}
  \caption[]{The dependence of the FOV averaged LECS dark Earth
             count rate on time. The launch date was 1996 April 30}
\label{fig:pb_total_lc}
\end{figure*}

A total of 498.6~ks of NXB data was accumulated using LECS dark Earth
pointings. 
The dependence of the 0.1--10~keV count rate, ${\rm C_T}$, (s$^{-1}$) 
on time is shown in Fig.~\ref{fig:pb_total_lc}, where each data point 
is averaged over the entire FOV. 
The interval without data points 
results from successive gyro failures on \sax,
which led to a 3 month observing hiatus. 
A gradual reduction in
the counting rate by $\sim$15\% over an interval of
$\sim$2~years is evident. This may be modeled as 
${\rm C_T = 0.1069 - 2.2 \times 10^{-5} \Delta T}$,
where ${\rm \Delta T}$ is the number of days since launch.
Alternatively, an exponential fit indicates a decay constant
of 11.8~years. The same trend is visible in data extracted
from the central 8\arcmin\ radius and background semi-annuli
regions (see Sect.~\ref{sect:bg_annuli}).
The intensity of the primary (5.89~keV Mn~K$_\alpha$ and 
6.49~keV Mn~K$_\beta$) radiation from the $^{55}$Fe calibration sources 
decays with a much shorter half-life of 2.73~years. However, the
calibration sources contain small amounts of other radioactive 
materials, the decay of which could contribute to the decrease in
NXB intensity. Another factor may be the decreasing particle background
in low-Earth orbit
as the next solar maximum is approached, due to the varying height of
the Earth's atmosphere. Such an effect is evident in studies of the ROSAT
High Resolution Imager NXB (\cite{snowden98}).
There is little evidence in the LECS data for 
``Long Term Enhancements'' or LTEs as seen by the PSPC (\cite{snowden95}).
These count rate increases are strongest below 0.28~keV and are approximately
uniform over an orbit. They last between 15--30 ROSAT orbits and
have peak intensities comparable to the low-energy CXB.

In the central 8\arcmin\ radius
extraction region the 0.1--10~keV NXB count rate is
$5.2 \times 10^{-6}$~s$^{-1}$~keV$^{-1}$~arcmin$^{-2}$. 
The dependence of the NXB on position within the FOV is
shown in the right panel of Fig.~\ref{fig:image}. 
The difference in
intensity between the most and least intense regions
is a factor 3.1. 
In contrast to the standard background (left panel of Fig.~\ref{fig:image}), 
the NXB intensity near the center of the FOV does not show 
an enhancement. Within the central 5\arcmin\ radius, any such
increase is $<$4\% of the total intensity.

\begin{figure}
  \centerline{\psfig{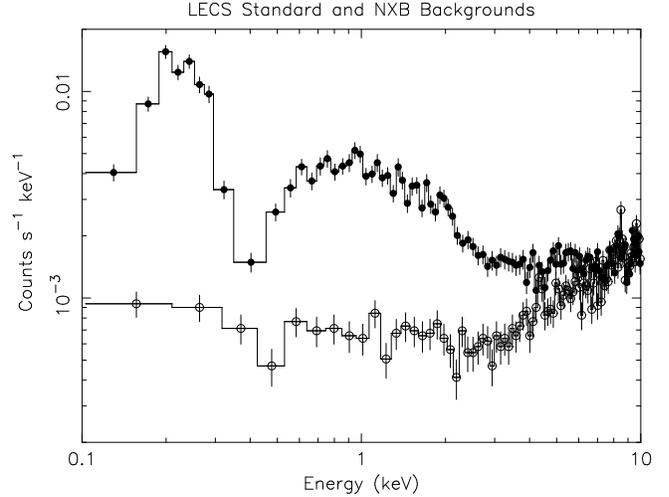}}
  \caption[]{Rebinned spectrum of the standard background (filled circles)
             and the NXB (open circles) in the central 8\arcmin\ radius
             of the FOV}
\label{fig:skybg_spec}
\end{figure}

\begin{figure}
  \centerline{\psfig{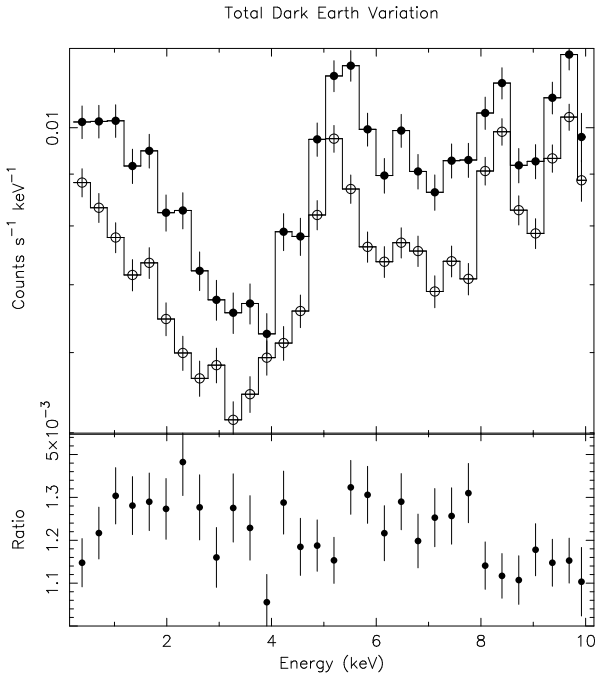}}
  \caption[]{The dependence of the rebinned LECS NXB spectrum, averaged 
             over the entire FOV, on time. The filled circles show the 
             spectrum obtained
             before day 400 and the open circles after day 400.
             The lower panel shows the ratio of the two spectra}
             \label{fig:pb_earlylate_spec}
\end{figure}

Spectra of the LECS NXB are shown in Figs.~\ref{fig:skybg_spec}
and \ref{fig:pb_earlylate_spec}.
The overall level is approximately
constant with energy with 3 discrete features superposed 
on a smooth increase $\approxgt$4~keV. These features
can be modeled as narrow Gaussian 
emission lines at $5.12 \pm 0.04$~keV (at 68\% confidence),
$8.51 \pm 0.04$~keV and $10.73 \pm 0.07$~keV. 
The first feature shows a clear position dependence, being stronger
close to the calibration sources. It is almost certainly produced 
by 5.9~keV characteristic X-rays from the 
$^{55}$Fe calibration sources
that penetrate deeply into the detector before being absorbed. The detected
energy is lower than the natural energy of the events,
because X-rays absorbed deep within the
detector produce, on average, less light than X-rays of the same 
energy absorbed close to the entrance window due to the different
scintillation lengths in the driftless gas cell (Parmar et al. 1997b).  
The other 2 features do not exhibit an obvious position dependence
within the FOV and may originate
from fluorescent excitation of L-shell transitions in the tungsten 
window support structure (see \cite{parmar97b}).

To illustrate the spectral changes associated with the decrease
in count rate shown in Fig.~\ref{fig:pb_total_lc}, 
Fig.~\ref{fig:pb_earlylate_spec} shows the spectrum of the NXB before
and after day 400. This shows that at energies $\approxlt$8~keV
the long term temporal evolution of the 
LECS NXB is not strongly energy dependent and may be simply modeled as
the change in overall normalization given above.
Above 8~keV, the intensity variation with time is less marked.
This may indicate that the fluorescent line features exhibit 
less intensity variability than the rest of the spectrum.

The geomagnetic rigidity is a measure of the minimum momentum required
by a cosmic particle to penetrate the Earth's magnetic field down to the
position of the satellite.
Due to the almost
circular, low inclination \sax\ orbit, the variation in rigidity
around the \sax\ orbit is less than is typical for
low-Earth orbiting spacecraft such as ASCA which has an orbital 
inclination of 31$\degmark$ and experiences rigidities between 
6--14~GeV~c$^{-1}$.
In the case of \sax, the rigidity typically varies 
from 10 to 16~GeV~c$^{-1}$ around the orbit. 
In order to investigate the dependence of the NXB spectrum on
rigidity, spectra were accumulated for intervals when the
rigidity was $\leq$ and $>$13~GeV~c$^{-1}$. The two spectra have almost
identical shapes, with the spectrum accumulated when the rigidity
was $\leq$13~GeV~c$^{-1}$ having an overall normalization $5 \pm 1$\%
higher. This means that for all but the shortest observations 
temporal averaging will ensure that the dependence of the NXB on 
geomagnetic rigidity can be ignored.


\begin{figure}
  \centerline{\psfig{figure=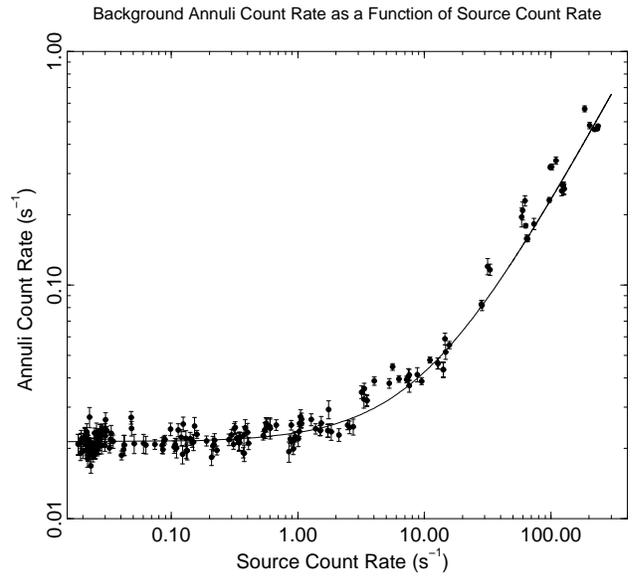,width=10.0cm,angle=0}}
  \caption[]{The dependence of the 0.1--10~keV background semi-annuli 
             count rate on the source count rate in the central 8\arcmin\ 
             radius extraction region. 
             The solid line shows the best-fit model discussed in the text}
\label{fig:annuli_source}
\end{figure}

\section{Background semi-annuli}
\label{sect:bg_annuli}

The same data as in Sect.~\ref{sect:nxb} were used 
to create a set of spectra extracted from within two 
semi-annuli located 17$'$ from the
center of the FOV at positions diametrically opposite to the
locations of the radioactive calibration sources (see Fig.~\ref{fig:cartoon}). 
The location and sizes of the semi-annuli were carefully chosen
to ensure that source ``spill-over'' is only significant for 
bright sources, while the correction for mirror vignetting
is kept as small as possible. 
The sum of the semi-annuli geometric
areas is equal to that of the standard 8$'$ radius source extraction region.
Counts within the semi-annuli consist of both particle and
internal background events (i.e., the NXB), sky background events, 
and any contribution from source photons that ``spill-over'' into
the semi-annuli.
The effects of ``spill-over'' are illustrated in 
Fig.~\ref{fig:annuli_source} where the count rate in
the semi-annuli, ${\rm C_{ann}}$, is plotted against 
source count rate in the central 8$'$ radius source extraction
region, ${\rm C_{src}}$. 
Extended sources such as supernova remnants and 
clusters of galaxies, observations where the
target source is offset in the FOV, and
pointings near the galactic center are excluded. 
The solid line shows the fit to 
the expected linear relation:
${\rm C_{ann} = 0.0212 + 0.00212\, C_{src}}$. 
This simple relation ignores the 
the energy dependence of ``spill-over''.
It indicates that ``spill-over'' adds $<$10\% to the semi-annuli background
for on-axis source count rates $<$1~s$^{-1}$.

\begin{figure}
  \centerline{\psfig{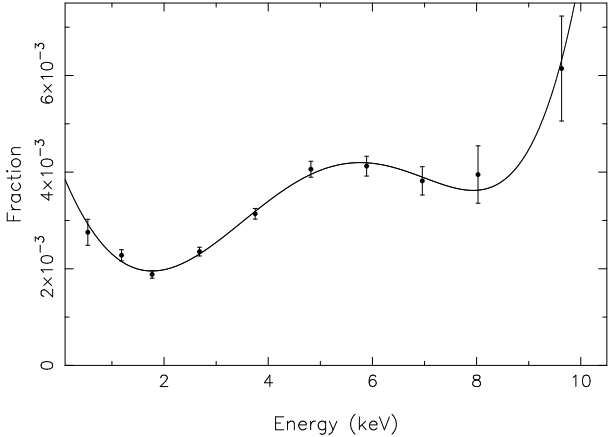}}
  \caption[]{Ratio of additional counts in the background semi-annuli
             divided by source counts as a function of energy. The
             solid line shows the fifth-order polynomial fit given in
             Table~\ref{tab:spilling_factor}}
\label{fig:scat_ratio}
\end{figure}

The energy dependence of ``spill-over'' was investigated by subtracting
the count rates obtained with blank fields from each of the semi-annuli 
spectra and dividing the remaining counts by those of the source with
the standard background subtracted.
Fig.~\ref{fig:scat_ratio} shows
a polynomial fit  (Table~\ref{tab:spilling_factor})
to the ratio derived in this way as a function of energy.
There are at least two effects which produce these additional counts in the
background semi-annuli. 
At low-energies ``spill-over'' is dominated by the poor detector 
spatial resolution (${\rm \propto Energy^{-0.5}}$), while
at high-energies the mirror scattering wings dominate. 

\begin{table}
\caption{Coefficients of the fifth-order polynomial fit to the
	     ratio of additional counts in the background semi-annuli
             divided by source counts as a function of energy in keV.
             These coefficients can be used to correct counts extracted from
             the background semi-annuli for source
	     contamination. a$_{0}$ is the constant term}
\begin{flushleft}
\begin{tabular}{lccc} 
\hline\noalign{\smallskip}
${\rm a_{n}}$ & \multicolumn{1}{c}{Value} &  ${\rm a_{n}}$ & 
\multicolumn{1}{c}{Value}\\
\noalign{\smallskip\hrule\smallskip}
$a_{0}$ & $-2.71\times10^{-3}$ & $a_{3}$ & $-6.13\times10^{-5}$ \\
$a_{1}$ & $+4.11\times10^{-3}$ & $a_{4}$ & $-1.01\times10^{-5}$ \\
$a_{2}$ & $+9.79\times10^{-4}$ & $a_{5}$ & $+9.57\times10^{-7}$ \\
\noalign{\smallskip}
\hline
\end{tabular}
\end{flushleft}
\label{tab:spilling_factor}
\end{table}

\begin{figure}
  \centerline{\psfig{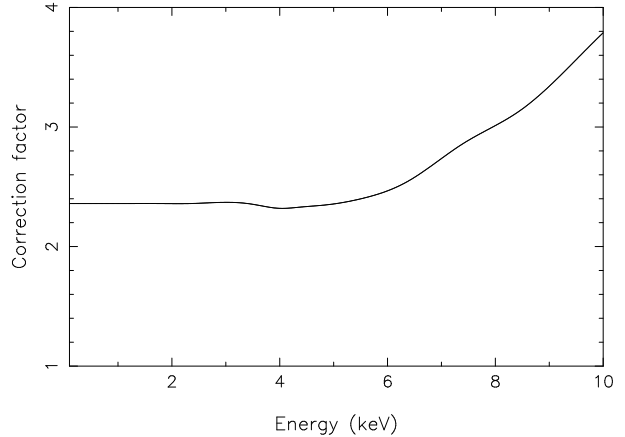}}
  \caption[]{Correction factors to be applied to a sky spectrum obtained
             from the background semi-annuli to give the predicted sky 
             background in the standard 8\arcmin\ radius
             source extraction region}
\label{fig:vign_spline}
\end{figure}

The blank field exposures listed in Table~\ref{tab:blank_fields} were 
also used to derive the offset correction factors which are shown 
in Fig.~\ref{fig:vign_spline}. 
Above 4~keV, where there are few sky background counts, the correction
factors were derived by extrapolating results obtained from a Crab Nebula
observation at an offset of 11\arcmin.
Multiplication of the 
NXB subtracted spectrum in the semi-annuli by these factors gives
the predicted background spectrum in the central 8\arcmin\ radius source
extraction region when the NXB is added. 
These correction factors are not
simply the mirror vignetting function given in Conti et al. (1994)
since, other factors such as an increase in PMT
noise due to the lower signal levels off-axis and the position 
dependency of singularly reflected X-rays may contribute. 
The correction factors
do not change significantly within the range of nominal target locations
within the FOV of $\sim$2\arcmin\ (see Sect.~\ref{sect:data_sel}).
Correction factors have also been determined for extraction radii
of 4\arcmin\ and 6\arcmin. These may be obtained by scaling the values
in Fig.~\ref{fig:vign_spline}, which are for an extraction radius of
8\arcmin, by the ratio of extraction region areas.
The ability to accurately derive these correction factors is limited
by the low number of counts in the standard background exposure.
As more LECS background fields become available, these correction
factors will be updated.

\section{Background subtraction techniques}
\label{sect:bgtech}

In order to study the X-ray emission from a particular source it
is usually necessary to subtract the X-ray and NXB contributions
from within the extraction region. In the case of the LECS (and MECS)
it is important to realize that strong sources outside the FOV (and
within 2$\degmark$ from the instrument axis) can contribute to
the background. A search should be made for the presence of such sources
using e.g., the HEASARC's XCOLL catalog.

In the case of the LECS, we present 3 ways of subtracting the
background. An estimate of the systematic uncertainty associated
with background subtraction may be obtained by comparing results obtained
using the different methods. 

\subsection{Standard background}

This is the simplest approach since the standard background includes
the contribution from the high ($>$$\vert 25 \vert \degmark$) galactic
latitude CXB and the NXB. Since the standard
background has a longer accumulation time (558.6~ks) than individual \sax\
observations, this method does not usually add significant noise.
Since the standard background varies with position within the 
FOV (Fig.~\ref{fig:image}), the background spectrum should 
be accumulated using the same extraction region as the source.
Note that this technique will underestimate the low-energy background
for pointings in directions with bright Galactic emission
such as the North Polar Spur (see Sect.~\ref{sect:standard_background}).

\subsection{Background semi-annuli}

This technique has the following steps:
\begin{enumerate}

\item{} Accumulate a spectrum of the two background semi-annuli 
during the observation of interest. These can be selected in RAWX, RAWY
using the data extraction program {\sc xselect} as follows:

\begin{verbatim}
 CIRCLE(128.0,128.0,80.0)
-CIRCLE(128.0,128.0,62.85)
-BOX(64.0,192.0,128.0,128.0)
-BOX(192.0,64.0,128.0,128.0)
\end{verbatim}
 
\item{} For source count rates $\approxgt$1~s$^{-1}$, 
subtract the NXB background
in the standard extraction region from the total counts in the same region.
Then multiply the subtracted counts by the function shown in 
Fig.~\ref{fig:scat_ratio} to derive the ``spill-over'' contribution.
This is then subtracted from the extracted semi-annuli spectrum.

\item{} Subtract the standard NXB spectrum of the semi-annuli.

\item{} Multiply the remaining spectrum
by the correction factors shown in Fig.~\ref{fig:vign_spline}. 
This provides an estimate of the CXB
in the source region. 

\item{} Add the standard NXB spectrum in the source region 
to this spectrum. This is then the final background to be subtracted 
before scientific analysis of the source properties.

\end{enumerate}

\subsection{Scaled ROSAT PSPC all-sky survey background}

ROSAT PSPC all-sky survey diffuse background maps are available 
in 7 energy bands with a pixel size of $12\arcmin \times 12$\arcmin\
covering $\sim$98\% of the sky.
The effects of discrete source counts, non X-ray contamination and 
X-rays of solar system 
origin have been eliminated to the greatest possible extent.
The contributions of all the sources listed in Table~\ref{tab:sources}
have been excluded.
These maps can be used to predict the variation
with pointing position of the LECS background.
The PSPC count rates closest to the centers of each of the fields 
included in the LECS standard background were first determined. 
The mean, LECS exposure weighted, PSPC 0.1--2.5~keV count rate
is $1.2 \times 10^{-3}$~s$^{-1}$~arcmin$^{-2}$. 

In order to understand how best to model the
PSPC data from different sky positions, 
7 channel PSPC spectra were determined for the fields given
in Tables~\ref{tab:blank_fields} and~\ref{tab:sources}.
These were then fit with a number of different
models including power-laws, thermal bremsstrahlungs, broken power-laws 
with the hard component $\alpha$ fixed at 1.4 (the value  
expected above 1~keV), and power-laws with $\alpha = 1.4$,
together with thermal components. In all cases low-energy absorption
was included. It was found that (1) the power-law and 
(2) the fixed power-law together with a thermal component give 
consistently the best fits for almost all the fields. 
These two models are therefore used in the
background analysis except for the Gal cent-3 field, where a
broken power-law, with hard index $\alpha$ fixed at 1.4 is preferred.

In order to use this technique to predict the LECS background spectrum
at a particular pointing position, the following steps are necessary:

\begin{enumerate}
\item Determine the ROSAT PSPC count rates in the 7 energy bands at the 
required position (see Sect.~\ref{sect:data_products}).

\item Fit spectral models to the derived spectrum
using the appropriate ROSAT all-sky survey response matrix. 

\item Select the model with the lowest reduced $\chi ^2$.

\item Use the standard LECS response matrix and the parameters
of the selected model fit to calculate a {\it predicted}
background spectrum for the position of interest.

\item Use the standard LECS response matrix and the CXB fit parameters
given below to calculate a {\it predicted} 
spectrum for the standard background field.

\item Divide the {\it predicted} LECS spectrum by that {\it predicted} for the
standard background to derive a set of PI channel dependent correction 
factors. 

\item Subtract the standard NXB spectrum (see Sect.~\ref{sect:nxb})
measured in the source extraction region from the
{\it observed} standard background spectrum.

\item Multiply the subtracted spectrum by the correction factors
calculated above.

\item Add the standard NXB measured in the source extraction region from the
standard background spectrum to the scaled spectrum.
This is then the final background to be subtracted 
before scientific analysis of the source properties.

\end{enumerate}

The LECS CXB spectrum is discussed in a companion {\it paper}
(\cite{parmar99}). In the 0.1--7.0~keV energy range it is 
well modeled by an absorbed power-law together with two 
solar abundance thermal bremsstrahlung components
(the {\sc mekal} model in {\sc xspec}). 
When this model is fit to the exposure weighted PSPC
spectrum corresponding to the standard background, the best-fit
results are a power-law $\alpha$ of 1.48 with normalization of 
$1.07 \times 10^{-3}$~photon~cm$^{-2}$~s$^{-1}$~keV$^{-1}$,
a hard bremsstrahlung
component temperature (kT) of 0.80~keV with normalization of 
$1.5 \times 10^{-5}$~photon~cm$^{-5}$, and a soft bremsstrahlung
component kT of 0.159~keV with normalization of 
$1.9 \times 10^{-4}$~photon~cm$^{-5}$. For both absorbed models, 
${\rm N_H}$ equivalent to $3.7 \times 10^{20}$~\hcm\ is required.
These fit parameters should be used to generate the {\it predicted}
standard LECS background spectrum.

\section{Background subtraction comparison}

The accuracy of the different background subtraction 
techniques described in Sect.~\ref{sect:bgtech}
was investigated using both blank field and exposures 
containing sources. The blank field exposures are listed in 
Table~\ref{tab:blank_fields} and the source exposures in
Table~\ref{tab:sources}. 

\subsection{Target selection}

CAL\,83 and CAL\,87 are two ``supersoft''
sources located in the Large Magellanic Cloud. CAL\,83 is only
detected in the LECS between 0.1--0.6~keV (\cite{parmar98}) and CAL\,87
between 0.2--1.0~keV (\cite{parmar97a}). 
HZ\,43 is a nearby hot white dwarf (e.g., \cite{barstow95})
that is often used as an extreme ultra violet or soft X-ray ``standard candle''
in X-ray astronomy. Due to the softness of its
spectrum, it is only detected below 0.28~keV in the LECS.
These 3 targets were chosen since they are only detected in a narrow
energy range of the LECS, allowing the rest of the spectrum to be used
to estimate the background subtraction quality.  

4U\,1630$-$47 is a recurrent X-ray transient located close to the
galactic plane. The \sax\ observation
was designed to detect quiescent emission between outbursts, but none
was found (\cite{oosterbroek98}). X\,1755$-$338
is a bright X-ray dipping source which was observed to have turned-off 
prior to the \sax\ observation (\cite{roberts96}). These 
two fields were chosen as tests of the background subtraction 
techniques in complex regions of the sky at low galactic latitudes.

NGC\,7172 is a Seyfert~2 galaxy, whose nuclear continuum is seen through
an ${\rm N_H \sim 10^{23}}$~\hcm\ neutral absorber. A soft excess observed
by ASCA was associated with diffuse emission from the group HGC\,90, in which
NGC\,7172 is located (\cite{guainazzi98}). 
In addition, two relatively bright ($\sim$0.25~count~s$^{-1}$) sources 
are included as a control sample, to show the relative independence of the
results on the choice of the background subtraction method at higher
flux levels. These are 
the Seyfert~2 Galaxy NGC~1068 (\cite{guainazzi99}) and the coronal
X-ray source VY\,Ari (\cite{favata97}). 

\begin{table*}
\caption[]{Observations used for background accuracy checks. CR is
LECS count rate including the contribution of the background in the
8\arcmin\ radius extraction regions}
\begin{flushleft}
\begin{tabular}{lllrrl}
\hline\noalign{\smallskip}
Source & CR & \hfil Date & \multicolumn{1}{c}{${\rm l_{II}}$} 
& \multicolumn{1}{c}{${\rm b_{II}}$} \hfil & \hfil Remarks\\
 Name  & (s$^{-1}$) & (yr~mn~day) & \multicolumn{1}{c}{$(\degmark)$} 
&\multicolumn{1}{c}{$(\degmark)$}\\
\noalign{\smallskip\hrule\smallskip}
CAL\,87             & 0.031  & 1996 Oct 27 & 281.8 & $-$30.7 & Supersoft source \\
4U\,1630$-$47 Field  & 0.040  & 1997 Mar 26 & 336.9 & +0.3  &X-ray transient
in quiescence   \\
CAL\,83             & 0.043  & 1997 Mar 07 & 278.6 & $-$31.3  & Supersoft source\\
X\,1755$-$338 Field & 0.045  & 1998 Apr 10 & 357.2 & $-$4.9  & X-ray binary in
quiescence \\
NGC\,7172           & 0.059  & 1996 Oct 14 & 321.2 & $-$18.6 &Active Galactic
Nuclei \\
NGC\,1068           & 0.142  & 1996 Dec 30 & 172.1 & $-$51.9 &Active Galactic
Nuclei \\
HZ\,43              & 0.215  & 1998 Jan 27 & 54.1  & +84.2 & Hot white dwarf   \\
VY\,Ari             & 0.303  & 1996 Sep 06 & 150.6 & $-$25.4 &Coronal X-ray 
emission \\
\noalign{\smallskip}
\hline
\end{tabular}
\end{flushleft}
\label{tab:sources}
\end{table*}

\begin{figure*}
\begin{center}
\psfig{figure=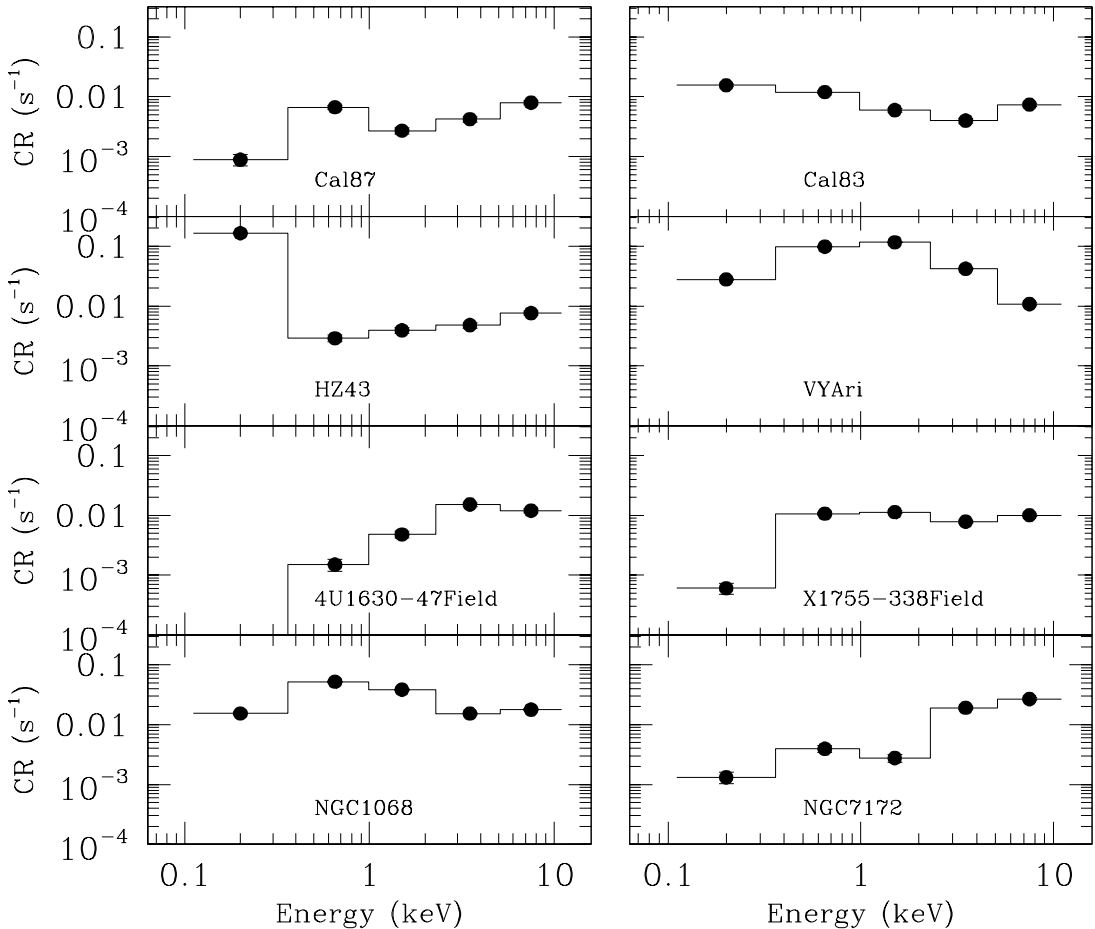,width=16.8cm}
\end{center}
\caption{Count rates (CR) for the source sample
in the energy ranges 0.1--0.3, 0.3--1, 1--2, 2--5, and 5--10~keV. 
The counts include contributions from the source (if present) and
any extended emission components within the 8\arcmin\ extraction
radii. The ordinate uncertainties are smaller than the symbol size}
\label{fig:figmg1}
\end{figure*}

\subsection{Results}

In Fig.~\ref{fig:figmg1}, observed LECS count rates in five energy bands
(0.1--0.3, 0.3--1, 1--2, 2--5, and 5--10~keV) are shown for the sample sources.
No correction for the instrumental response has been applied.
All results in this section
are displayed in this way to allow ease of comparison since
count rate
is a quantity which can be easily estimated when comparing the behavior
of a weak X-ray source with the instrumental performances and limitations.

\begin{figure*}
\begin{center}
\psfig{figure=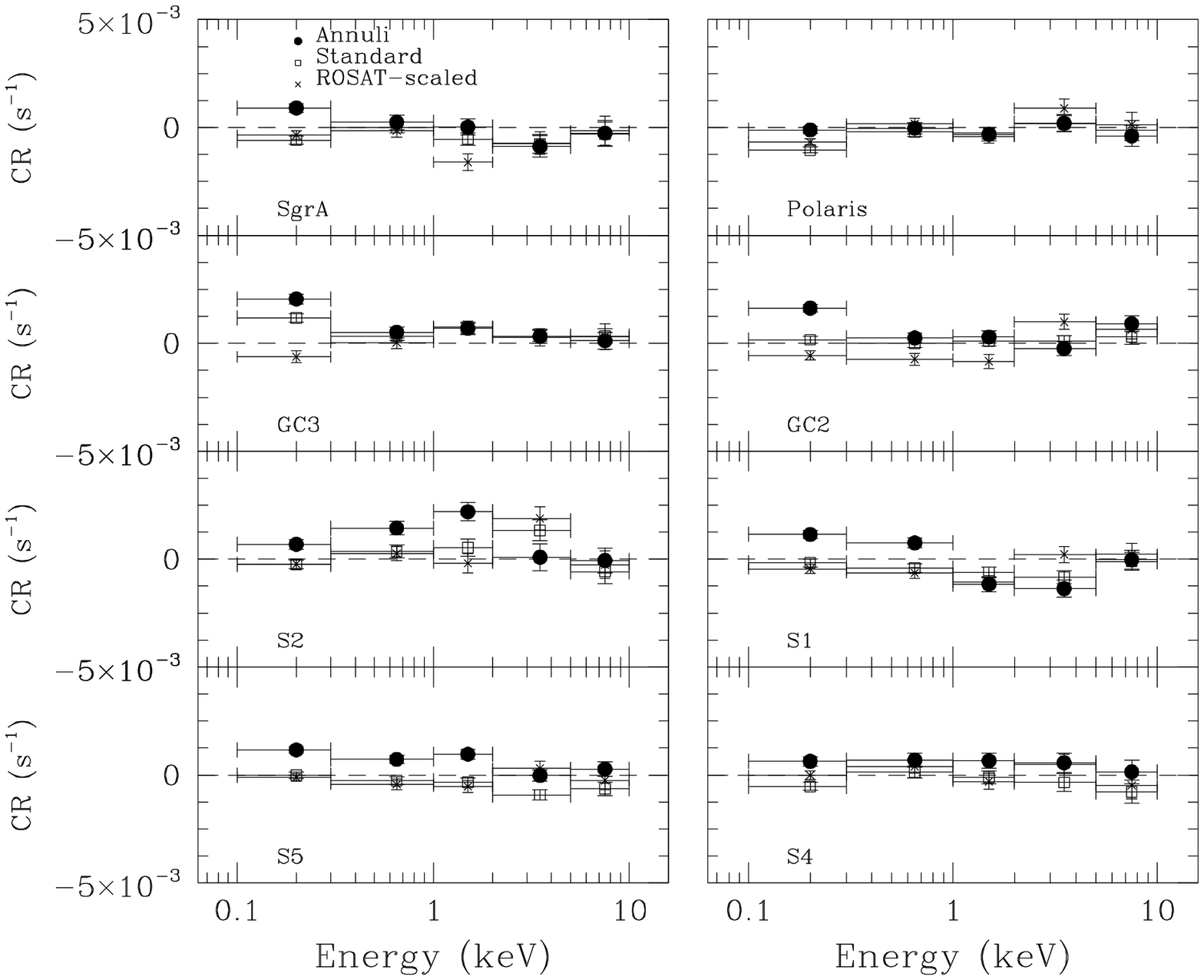,width=16.8cm}
\end{center}
\caption{Count rate residuals (CR) as a function of energy for 8 blank
fields from Table~\ref{tab:blank_fields}, when the 3 background
subtracted techniques are applied. 
Empty squares indicate the standard background,
filled circles the semi-annuli, and crosses the ROSAT-scaled techniques}
\label{fig:figmg2}
\end{figure*}

\begin{figure}
\begin{center}
\psfig{figure=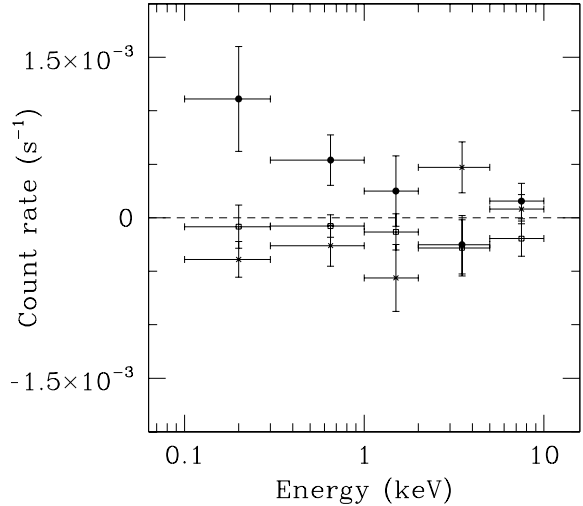,width=8.0cm}
\end{center}
\caption{Exposure time weighted sum of the 8 blank field exposure
residuals when the 3 background subtraction techniques are
applied. Empty squares indicate the standard background,
filled circles the semi-annuli, and crosses the ROSAT-scaled techniques}
\label{fig:figmg3}
\end{figure}

Fig.~\ref{fig:figmg2} shows the residuals 
for 8 blank fields when each of the 3 background subtraction
techniques are applied. (The shortest exposure blank field, 
the SDC Target, in
Table~\ref{tab:blank_fields} was excluded.)
In the case of a good background subtraction, the residuals should exhibit a
Poissonian distribution centered on zero, and the standard deviation of the
residuals provides an estimate of the
systematic uncertainties associated with each technique.
These results are summarized in Fig.~\ref{fig:figmg3}, where the
exposure time weighted sum of the
background field residuals, shown individually 
in Fig.~\ref{fig:figmg2}, are plotted. 
The average mean count rates, ${\rm \mu}$,
in the 0.1--10~keV energy range are
$(-1.5 \pm 0.9)$, $(3.6 \pm 1.6)$ and
$(-1.3 \pm 1.0) \times 10^{-4}$~s$^{-1}$, for the standard, semi-annuli
and ROSAT-scaled technique, respectively.
Table~\ref{tab:summary} lists
the values of ${\rm |\mu| + 3 \sigma}$
(${\rm \sigma}$ is the standard deviation) for the
3 techniques in 3 energy ranges.
These values may be interpreted
as 3$\sigma$ estimates of the systematic uncertainties
of each the background subtraction techniques at high galactic latitudes. 
In the 0.1--2~keV energy range, the
values in Table~\ref{tab:summary} correspond to fluxes in the
range 0.7--1.8 and 0.5--$1.1 \times 10^{-13}$~erg~cm$^{-2}$~s$^{-1}$
for a power-law spectrum with $\alpha = 2.0$
and photoelectric absorption, ${\rm N_H}$, of ${\rm 3 \times
10^{20}}$ and ${\rm 3 \times 10^{21}}$~atom~cm$^{-2}$, respectively.
In the 2--10~keV band, for an \nh\ of ${\rm 3 \times
10^{20}}$~atom~cm$^{-2}$, the limiting fluxes are 2--3 and
1.3--$3 \times 10^{-13}$~erg~cm$^{-2}$~s$^{-1}$ for sources
with $\alpha = 2.0$ and 1.5, respectively.

In the case of the blank fields, the standard background provides the
most consistent (i.e., lowest values of ${\rm |\mu| + 3 \sigma}$) subtraction.
This is unsurprising given that the standard background {\it itself} includes
the comparison fields (see Table~\ref{tab:blank_fields}). In addition,
the blank field exposures are at high ($>$$\vert 25 \vert \degmark$) 
galactic latitude and avoid features such as the North Polar Spur,
and so are expected to be broadly similar. The residuals obtained
with the semi-annuli method are the largest. An investigation 
reveals that this may be in part due to uncertainties in 
the NXB spectra of the two semi-annuli which contain
$<$8800 counts. It is
expected that this will improve as subsequent dark Earth exposures 
are added to the already existing data set. The ROSAT-scaled background
works consistently well within the limits specified in Table~\ref{tab:summary}.

\begin{figure*}
\begin{center}
\psfig{figure=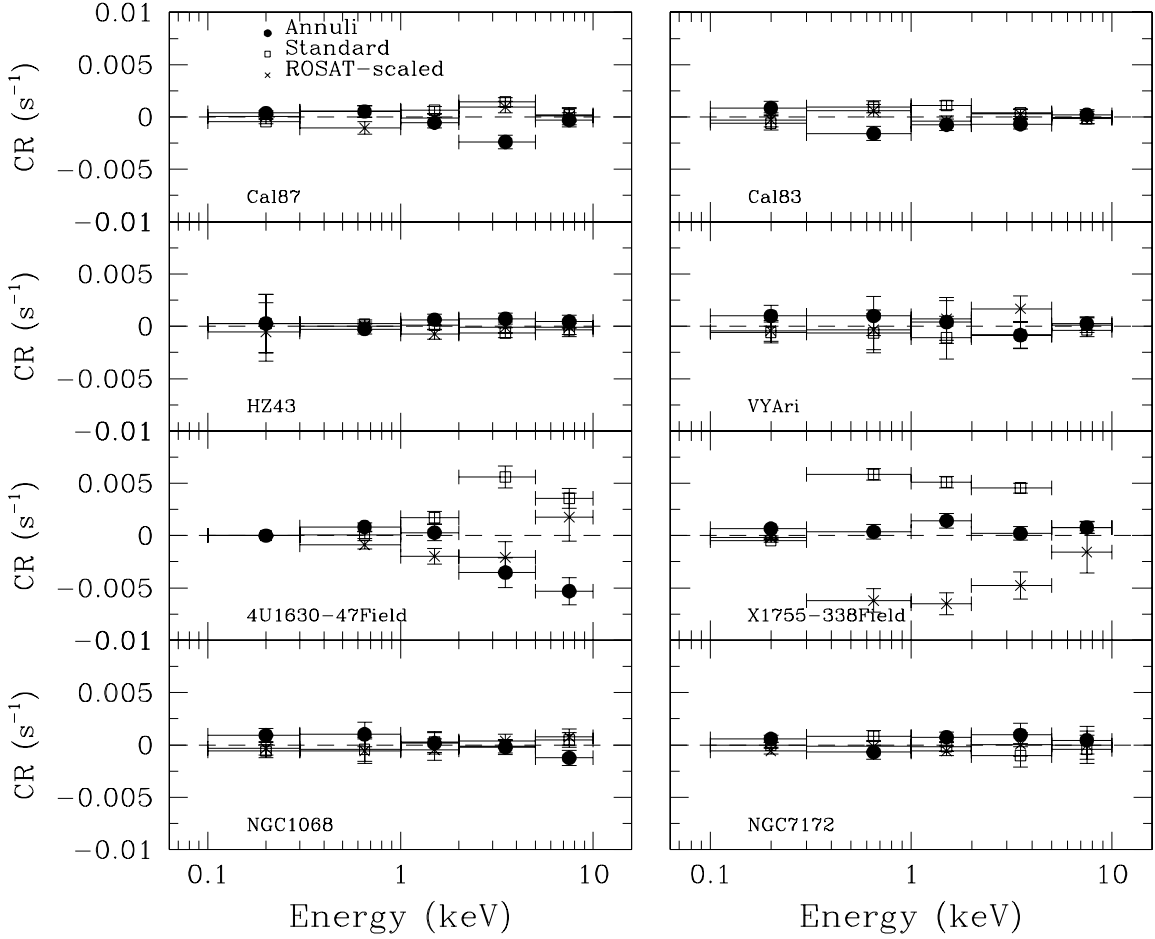,width=16.8cm}
\end{center}
\caption{Count rate difference compared to the mean (CR) as a
function of energy when the 3 background subtraction techniques are
applied to each of the sample sources. 
Empty squares indicate the standard background technique,
filled circles the semi-annuli, and crosses the ROSAT-scaled technique.
Note that the ordinate extrema are a factor 2 larger than in 
Fig.~\ref{fig:figmg2}}
\label{fig:figmg4}
\end{figure*}

Fig.~\ref{fig:figmg4} illustrates the application of the 3
background subtracted techniques to the sample sources. For each
source the 3 techniques were applied and background subtracted
count rates in the 5 energy bands calculated. The difference with
respect to the mean of the 3 techniques is plotted in 
Fig.~\ref{fig:figmg4}.
In 6 out of 8 cases no significant deviations at levels higher
than a few $\times 10^{-3}$~count~s$^{-1}$ are present.
This is comparable to the results on the blank sky fields and indicates
that all three methods work well here. The differences between the three
methods may be used to estimate the systematic uncertainties associated
with background subtraction.

In the case of the two
fields close to the galactic plane (4U\,1630$-$47 and X\,1755$-$338), 
differences at a level $\approxlt$$10^{-3}$~count~s$^{-1}$ are evident. 
This is a factor $\sim$2.5 larger than with the blank field exposures
and is probably due to incorrect
estimation of the contributions of the hard diffuse emission associated with
the Galactic ridge, or unresolved point-sources. In addition, in the
case of 4U\,1630$-$47 the 50~mCrab (or $\sim$10 LECS count~s$^{-1}$) 
X-ray binary 4U\,1624$-$49 is located
2$\degmark$ away. X-rays from 4U\,1624$-$49 that undergo a single mirror 
reflection may provide a small contribution to the 4U\,1630$-$47 field.

In both these cases, the standard background systematically underestimates 
the background (as expected), and 
the scaled ROSAT method overestimates the
background somewhat, when compared to the mean values.
The reason for this is unclear, but it may be partly due
to incorrect ROSAT source subtraction in the complex fields.
In the case of the X\,1755$-$338 field, 
the residuals obtained using the semi-annuli method are
smoothly distributed between those obtained using
the other two methods. This is not the case in the 4U\,1630$-$47 field
where the residuals deviate from the mean values above 2~keV.
These results suggest that the semi-annuli
gives the most reliable results for complex fields, with the 
scaled ROSAT method being the second most reliable.  
These two cases 
illustrate the difficulty in obtaining a good background
subtraction for sources located at low galactic latitudes, or in complex
regions of the X-ray sky. 

\begin{table}
\caption{The values of ${\rm |\mu| + 3 \sigma}$ for the 3
background subtraction techniques in 3 energy ranges for the blank fields in
units of $10^{-3}$~count~s$^{-1}$}
\begin{flushleft}
\begin{tabular}{lccc} 
\hline\noalign{\smallskip}
Method & 0.1--10~keV & 0.1--2~keV & 2--10~keV \\
\noalign{\smallskip\hrule\smallskip}
Standard     & 0.8 & 0.7 & 1.3 \\
Semi-annuli  & 1.4 & 1.9 & 1.2 \\
ROSAT-scaled & 0.9 & 1.3 & 1.3 \\ 
\noalign{\smallskip}
\hline
\end{tabular}
\end{flushleft}
\label{tab:summary}
\end{table}

Finally, the effects of the long term decrease in 
NXB intensity evident in Fig.~\ref{fig:pb_total_lc} were evaluated using
all three proposed background subtraction techniques. In all cases
the differences in background residuals were substantially smaller
than the values given in Table~\ref{tab:summary}. This means that 
currently no time dependent corrections need to be applied
to the LECS background.

\section{Data products}
\label{sect:data_products}

The standard background and NXB cleaned event lists will be made
available by the \sax\ Science Data Center in Rome.
These files will allow users to generate appropriate
background and NXB spectra.

In addition, the Max-Planck-Institut f\"ur Extraterrestrische Physik, 
in Garching will make the 7 energy channel source subtracted PSPC 
maps available
through the World Wide Web so that users can use the scaled ROSAT
background subtraction technique.

\begin{acknowledgements}
The \sax\ satellite is a joint Italian-Dutch programme.
M. Guainazzi, T. Oosterbroek, and A. Orr acknowledge ESA Fellowships
and N. Shane the ESA stagiaire programme. 
We thank F. Fiore for pointing out the importance of single 
reflection X-rays to this study, and S. Molendi and S. Snowden 
for helpful discussions. The support of M. Werger
and O.R. Williams in maintaining the SSD \sax\ archive is appreciated.
Finally, we thank all the \sax\ Principal Investigators who allowed
their data to be used for this study prior to them becoming public.
\end{acknowledgements}

\end{document}